\documentclass[aps,prc,twocolumn,groupedaddress,showpacs]{revtex4}

\usepackage{graphicx}
\usepackage{dcolumn}
\usepackage{bm}
\usepackage{longtable}

\begin{document}

\title{
The $^{15}$N($\bm\alpha$,$\bm\gamma$)$^{19}$F reaction and
nucleosynthesis of $^{19}$F
}

\author{S.\ Wilmes}
\altaffiliation{Allianz Vers.\ AG, K\"oniginstra{\ss}e 28, D-80802 M\"unchen,
Germany}
\author{V.\ Wilmes}
\altaffiliation{M\"adchenrealschule Weichs, Freiherrnstra{\ss}e 17, D-85258
Weichs, Germany}
\author{G.\ Staudt}
\affiliation{
Physikalisches Institut, Universit\"at T\"ubingen, 
Auf der Morgenstelle 14, D-72076 T\"ubingen, Germany
}

\author{P.\ Mohr}
\thanks{Corresponding Author}
\email[E-mail: ]{mohr@ikp.tu-darmstadt.de}
\affiliation{
Institut f\"ur Kernphysik, Technische Universit\"at Darmstadt,
Schlossgartenstra{\ss}e 9, D-64289 Darmstadt, Germany
}

\author{J.\ W.\ Hammer}
\affiliation{
Institut f\"ur Strahlenphysik, Universit\"at Stuttgart,
Allmandring 3, D-70569 Stuttgart, Germany
} 

\date{\today}

\begin{abstract}
Several resonances in the $^{15}$N($\alpha$,$\gamma$)$^{19}$F reaction 
have been investigated in the energy range between 0.6\,MeV and
2.7\,MeV. Resonance strengths and branching ratios have been
determined. High sensitivity could be obtained by the combination of
the {\sc{dynamitron}} high current accelerator, the windowless gas
target system {\sc{rhinoceros}}, and actively shielded germanium
detectors. Two levels of $^{19}$F could be observed for the first time
in the ($\alpha$,$\gamma$) channel, and several weak branchings below
the detection limits of previous experiments were measured. Two
observed resonances correspond to $\alpha$-cluster states in $^{19}$F
which have been assigned unambiguously. The astrophysical reaction
rate is derived from this set of resonance strengths.
\end{abstract}


\maketitle

\section{\label{sec:intro}Introduction}
Light nuclei are mainly synthesized by fusion reactions between
charged particles. Obvious exceptions are $^{6,7}$Li, $^{9}$Be, and
$^{10,11}$B, which are bypassed by the triple-alpha process
$3\,\alpha$ $\rightarrow$ $^{12}$C and are thus not produced in
stellar hydrogen or helium burning. The nucleus $^{19}$F may be
produced in a sidepath of the CNO cycle. However, because of the large
cross section of the $^{19}$F(p,$\alpha$)$^{16}$O reaction, the
freshly synthesized $^{19}$F is destroyed rapidly in any proton-rich
environment. As an alternative, the reaction
$^{15}$N($\alpha$,$\gamma$)$^{19}$F has been suggested as the main
production reaction, and the astrophysical scenario is assumed to be a
thermally pulsing AGB star
\cite{1992A&A...261..157F,1992A&A...261..164J}, and also Wolf-Rayet
stars have been suggested \cite{2000A&A...355..176M}.
A further possible
mechanism for $^{19}$F production is neutrino-induced
nucleosynthesis by $^{20}$Ne($\nu$,$\nu'$N) reactions in supernovae
\cite{1990ApJ...356..272W}.

The dominating reaction path for the production of $^{19}$F in AGB
stars is assumed to be
$^{14}$N($\alpha$,$\gamma$)$^{18}$F($\beta^+$)$^{18}$O(p,$\alpha$)$^{15}$N($\alpha$,$\gamma$)$^{19}$F
\cite{1992A&A...261..157F,1992A&A...261..164J}. Destruction of
$^{19}$F in a hydrogen-poor environment may proceed via
$^{19}$F($\alpha$,p)$^{22}$Ne which has been studied recently
\cite{Gor2002}. Typical temperatures are of the order of $T_9 = 0.2 -
0.3$ (where $T_9$ is the temperature in $10^9$\,K) which corresponds to 
effective energies around $350 - 500$ keV.

Additional astrophysical interest in the
$^{15}$N($\alpha$,$\gamma$)$^{19}$F reaction comes from the fact that
properties of the $^{19}$Ne mirror nucleus can be derived assuming
similar reduced widths $\theta_\alpha^2$ and radiation widths
$\Gamma_\gamma$. The $^{15}$O($\alpha$,$\gamma$)$^{19}$Ne reaction is the
bottleneck for the outbreak from the hot CNO cycle to the rapid proton
capture process \cite{1981ApJS...45..389W,1986ApJ...301..629L}, and
therefore reduced widths $\theta_\alpha^2$ have to be determined for
the first states above the $\alpha$ threshold in
$^{19}$Ne. Additionally, higher-lying resonances in
$^{15}$N($\alpha$,$\gamma$)$^{19}$F have been studied \cite{Butt98}
which correspond to the lowest resonances in the 
$^{18}$F(p,$\alpha$)$^{15}$O and $^{18}$F(p,$\gamma$)$^{19}$Ne
reactions.
Any direct experimental study of the
latter reactions is hampered by the unstable nuclei $^{15}$O and $^{18}$F.
However, there are significant uncertainties in the translation of
properties of $^{19}$F to its mirror nucleus $^{19}$Ne 
which may lead to uncertainties up to more than one order of magnitude
for the corresponding reaction rates \cite{Oli97}.

The astrophysical reaction rate of $^{15}$N($\alpha$,$\gamma$)$^{19}$F
is dominated by resonance contributions of several low-lying states in
$^{19}$F. Several ($\alpha$,$\gamma$) experiments on individual resonances have
been performed, but no previous experiment has covered a broad energy
range from several hundred keV up to a few MeV
\cite{Mag87,Rog72a,Rog72b,Dix71a,Dix71b,Ait70,Pri57}.
The available information is summarized in \cite{Ajz87,Til95,NACRE}.
This paper presents new experimental data in the energy range from
0.6\,MeV to 2.7\,MeV. All data have been measured with the same
set-up; this leads to a complete and consistent set of resonance
strengths. With the exception of the $7/2^+$ resonance at 461\,keV all
astrophysically relevant resonance strengths could be determined. The
strength of this resonance was measured indirectly in an
$\alpha$-transfer experiment using the $^{15}$N($^{7}$Li,t)$^{19}$F
reaction \cite{Oli96}. The
lowest resonance with $J^\pi = 9/2^-$ at $E_x = 4033$\,keV 
($E_\alpha = 24$\,keV) does not
contribute significantly to the reaction rate at the given
temperatures.

The paper is organized as follows: In Sect.~\ref{sec:exp} we present
briefly our experimental set-up. Results for the resonance strengths
and the branching ratios are given in
Sect.~\ref{sec:res}. $\alpha$-cluster states in $^{19}$F are discussed
in Sect.~\ref{sec:clu}. The astrophysical reaction rate is calculated
in Sect.~\ref{sec:ast}, and finally conclusions are drawn in
Sect.~\ref{sec:con}. 

In this paper energies $E_\alpha$ are given in the laboratory
system. The center-of-mass energy $E_{\rm{c.m.}}$ is related to
$E_\alpha$ and to the excitation energy $E_x$ by the well-known
formulae $E_{\rm{c.m.}} = E_\alpha \, \times \, A_T/(A_P+A_T) \approx
0.7894 \, E_\alpha$ and $E_x =
E_{\rm{c.m.}} + Q$ with $Q = 4013.8$\,keV.

\section{\label{sec:exp}Experimental Set-up}
The experiment has been performed at the windowless gas target system
{\sc{rhinoceros}} \cite{Hammer99}
which is installed at the {\sc{dynamitron}} accelerator
of Universit\"at Stuttgart. Details of the experimental set-up and the
data analysis are given in \cite{Koe99,Wilmes96}. Here we repeat briefly the
relevant properties of this set-up.

The $^{4}$He$^{+}$ beam with currents up to 120\,$\mu$A was focussed
into an extended windowless gas target area with a length of about
6\,cm. The N$_2$ target gas was highly enriched in $^{15}$N to 98\,\%
and 99.7\,\%. The gas was recirculated and purified by a cooling trap.
Typical target pressures were about 1\,mbar. Two diaphragms
with a length of 15\,mm and a diameter of 5\,mm were used to collimate
the $\alpha$ beam and to define the target area. Several additional
diaphragms had to be used to reduce the pressure by 8 orders of
magnitude to the vacuum in the beam transport system of $10^{-8}$\,mbar.
The windowless gas target system has the advantages of high stability
(it cannot be damaged by the beam), of high purity and of variable
density which can be easily changed by the target pressure.

The $\alpha$-beam was monitored by two silicon surface-barrier
detectors mounted either at 27.5$^\circ$, 60$^\circ$, or 90$^\circ$
relative to the $\alpha$ beam axis. The positions of the particle
detectors were chosen to obtain small deviations from the Rutherford
cross section at reasonable counting rates.

For the detection of the $\gamma$-rays we used two large-volume
high-purity germanium (HPGe) detectors with efficiencies up to about
100\,\% (relative to a $3'' \times 3''$ NaI(Tl) detector). One detector
was actively shielded by a BGO detector. The efficiency determination
of the HPGe detectors was obtained from GEANT simulations which were
shown to be reliable in the analyzed energy region \cite{Koe99}.
For a reliable determination of weak branching ratios we have taken
into account summing effects from cascade $\gamma$-rays.

\begin{figure}[hbt]
\includegraphics[ bb = 78 389 486 564, width = 85 mm, clip]{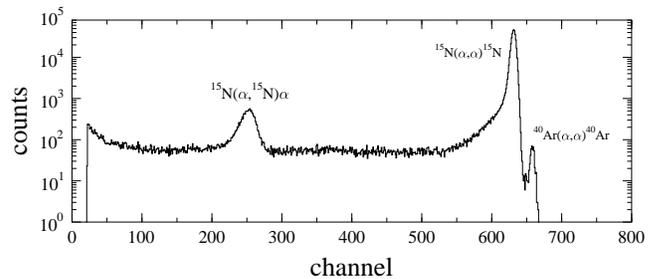}
\caption{
\label{fig:spec_alpha} 
Energy spectrum of the particle detector mounted at
$\vartheta_{\rm{lab}} = 27.5^\circ$. Besides the dominating peak from
$^{15}$N($\alpha$,$\alpha$)$^{15}$N also the recoil nucleus $^{15}$N
is clearly visible. Additionally, a weak contamination of the target
gas ($^{40}$Ar; $\approx 0.015\,\%$)
can be seen. The spectrum has been measured at $E_\alpha = 1640$\,keV.
}
\end{figure}

\begin{figure}[hbt]
\includegraphics[ bb = 156 144 562 385, width = 85 mm, clip]{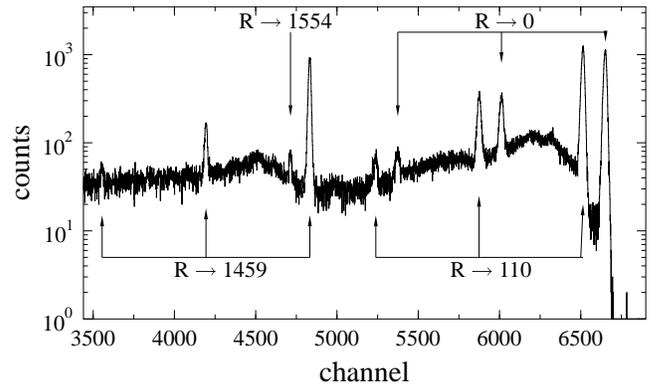}
\caption{
\label{fig:spec_gamma} 
Energy spectrum of the HPGe detector, measured at the $1/2^+$
resonance at $E_x = 5337$\,keV ($E_\alpha = 1676$\,keV). Transistions
to the final states at $E_x = 0$, 110, 1459, and 1554\,keV can be
seen. The corresponding full energy, single escape, and double escape
peaks are marked. This spectrum has
been measured in anticoincidence with the signal from the active BGO
shielding.
}
\end{figure}

Typical spectra are shown in Figs.~\ref{fig:spec_alpha} and
\ref{fig:spec_gamma}. Fig.~\ref{fig:spec_alpha} shows a spectrum of
the silicon detector mounted at $\vartheta_{\rm{lab}} = 27.5^\circ$,
and Fig.~\ref{fig:spec_gamma} presents a $\gamma$-ray spectrum of the
HPGe detector in the $1/2^+$ resonance at $E_x = 5337$\,keV.

\section{\label{sec:res}Resonance Strengths and Branching Ratios}
The resonance strength $\omega \gamma$ of a resonance is defined by
\begin{equation}
\omega \gamma = \omega \, \frac{\Gamma_\alpha \,
\Gamma_\gamma}{\Gamma}
\label{eq;strength}
\end{equation}
with a statistical factor $\omega = (2J+1) / [(2J_P+1) (2J_T+1)] =
(2J+1)/2$ for $^{15}$N($\alpha$,$\gamma$)$^{19}$F and the partial
widths $\Gamma_\alpha$, $\Gamma_\gamma$, and the total width $\Gamma =
\Gamma_\alpha + \Gamma_\gamma$ for excitation energies below other
particle thresholds of $^{19}$F ($S_{\rm{n}} = 10432$\,keV,
$S_{\rm{p}} = 7994$\,keV). The resonance strength is approximately
given by $\omega \gamma \approx \omega \Gamma_\alpha$ for
$\Gamma_\alpha \ll \Gamma_\gamma$ and $\omega \gamma \approx \omega
\Gamma_\gamma$ for $\Gamma_\gamma \ll \Gamma_\alpha$. The first case
is expected for resonances close above the $\alpha$ threshold at
4014\,keV whereas the second case can be found at higher energies.
For several states the ratio $\Gamma_\gamma/\Gamma$ has been measured
where the $^{12}$C($^{11}$B,$\alpha$)$^{19}$F reaction has been used
for the population of these states \cite{Pri89}.

\begin{figure}[hbt]
\includegraphics[ bb = 49 84 507 570, width = 85 mm, clip]{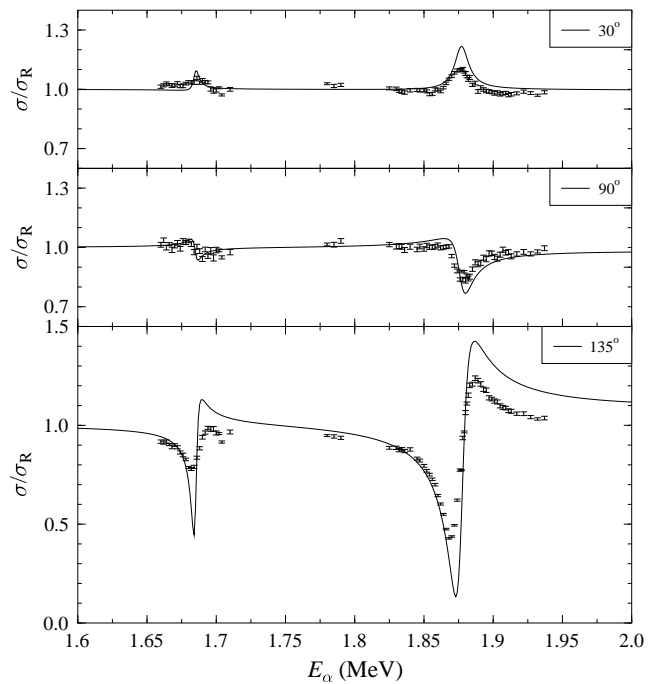}
\caption{
\label{fig:elast} 
Elastic scattering cross sections around the $J^\pi = 1/2^+$, $E_x =
5337$\,keV and $J^\pi = 3/2^+$, $E_x = 5501$\,keV resonances,
normalized to Rutherford cross 
section, measured with silicon detectors at $\vartheta_{\rm{lab}} =
30^\circ$ (upper diagram), $60^\circ$ (middle), and $135^\circ$
(lower). The full line is an optical model calculation (see
Sect.~\ref{sec:clu}).
}
\end{figure}

The resonance strength $\omega \gamma$ is related to the experimental
yield $Y$ by
\begin{equation}
Y = \frac{N_P \rho_T}{S} \frac{A_P + A_T}{A_T} 
\frac{\pi^2 \hbar^2}{\mu E_{\rm{c.m.}}} (\omega \gamma)
\label{eq:yield}
\end{equation}
with the number of projectiles $N_P$, the density of target atoms
$\rho_T$ (in cm$^{-3}$), the stopping power $S$ (in keV/$\mu$m), and
the reduced mass $\mu$. 
Resonance strengths and branching ratios for all measured resonances
have been determined by the following procedure: The product of the
number of projectiles and target atoms per cm$^2$ was derived from the
elastically scattered $\alpha$ particles in the silicon detectors. 
Using the geometry of the particle detectors, the product of the
number of projectiles $N_P$ and the target density $\rho_T$ is
calculated. The
elastic cross section was assumed to follow Rutherford's law for
pointlike charges. This assumption was tested by an additional
measurement using a different target chamber with three silicon
detectors at $\vartheta_{\rm{lab}} = 
30^\circ$, $60^\circ$, and $135^\circ$, and a mixture of xenon and
$^{15}$N as target gas. As 
can be seen from Fig.~\ref{fig:elast}, the deviations for the most
forward detector are very small and remain within about 10\,\% even
for the broad resonances at $E_\alpha = 1676$, 1884, and 2627\,keV. 
These deviations have been taken into account for the determination of
the strengths of the mentioned resonances. The stopping power $S$ was
taken from \cite{Zie77}. The energies of the resonances were taken
from \cite{Til95}; all energies were confirmed within their
uncertainties by the measured $\gamma$-ray energies.

The number of reactions was determined from the number of counts
in the peaks of the HPGe $\gamma$-ray spectra. The efficiency of the
HPGe detectors was calculated by GEANT simulations which were
verified by measurements with radioactive sources and by the analysis
of the $^{27}$Al(p,$\gamma$)$^{28}$Si reaction \cite{Koe99}.
The angular distribution of the $\gamma$-rays was taken into account.
We assumed pure $E1$ transitions for $\Delta I = 0, \pm 1$, $\pi =
-1$; for $\Delta I = 0, \pm 1$, $\pi = +1$ we calculated $M1/E2$
angular distributions with a mixing parameter $\delta$ from
\cite{Til95} whenever available or $\delta = 0$ otherwise; for $\Delta
I = 2$, $\pi = +1$ pure $E2$ radiation is assumed.
The close geometry of the HPGe detector with a distance of only 6\,cm
leads to a large coverage of $60^\circ \le \vartheta_\gamma \le
120^\circ$. Because of the half-integer spin of $^{15}$N the
deviations from an isotropic angular distribution are much smaller
compared to spin-zero target nuclei, e.g.\ in the reaction
$^{20}$Ne($\alpha$,$\gamma$)$^{24}$Mg, see \cite{Koe99}. Therefore,
the uncertainties remain small even in cases where $\delta$ is
unknown. 

The overall uncertainties are dominated by the efficiency
determination ($\alt 4\,\%$), the density profile of the target gas
(mainly for broad resonances, $\alt 5\,\%$), the stopping power ($\alt
5\,\%$), in rare cases by the
unknown mixing parameter $(\alt 10\,\%$), and sometimes by poor
statistics. The sizes of these uncertainties change from transition to
transition. The given values are typical leading to overall
uncertainties of the resonance strengths of $\alt 10\,\%$ and
much smaller uncertainties of the branching ratios.

The results for the resonance strengths are compared to literature
values in Table \ref{tab:strength}. The measured branching ratios are
listed in Table \ref{tab:branch}. In the following paragraphs further
information is given on all resonances.

\begin{table}[hbt]
\caption{
\label{tab:strength}
Resonance strengths $\omega \gamma$ for the resonances between
$E_\alpha = 461$ and 2642\,keV ($4378\,{\rm{keV}} \le E_x \le
6100\,{\rm{keV}}$). The results show overall good agreement with the
literature data $\omega \gamma_{\rm{Lit}}$ \cite{Til95}; detailed
references are given in the last column. All energies $E_\alpha$ and
$E_x$ are given in keV.
}
\begin{ruledtabular}
\begin{tabular}{rrcr@{$\pm$}lr@{$\pm$}lc}
$E_\alpha$
& $E_x$ 
& $J^\pi$
& \multicolumn{2}{c}{$\omega \gamma$ \footnote{this work}}
& \multicolumn{2}{c}{$\omega \gamma_{\rm{Lit}}$ \footnote{Ref.~\cite{Til95}}}
& Ref. \\
\hline
461	& 4378	& $7/2^+$
	& \multicolumn{2}{c}{$-$}
	& \multicolumn{2}{c}{$6^{+6}_{-3}$\,neV \footnote{from
$^{15}$N($^{7}$Li,t)$^{19}$F transfer experiment \cite{Oli96}}}
	& \cite{Oli96} \\
679	& 4550	& $5/2^+$
	& (95.5	& 11.7)\,$\mu$eV	
	& (97	& 20)\,$\mu$eV
	& \cite{Mag87} \\
687	& 4556	& $3/2^-$
	& (6.4	& 2.5)\,$\mu$eV	
	& \multicolumn{2}{c}{$< 10$\,$\mu$eV}
	& \cite{Mag87} \\
847	& 4683	& $5/2^-$
	& (5.6	& 0.6)\,meV	
	& (6	& 1)\,meV
	& \cite{Rog72a} \\
1384	& 5107	& $5/2^+$
	& (9.7	& 1.6)\,meV	
	& (13	& 8)\,meV
	& \cite{Ait70,Ajz72} \\
1676	& 5337	& $1/2^+$
	& (1.69	& 0.14)\,eV	
	& (1.64	& 0.16)\,eV
	& \cite{Dix71a} \\
1778	& 5418	& $7/2^-$
	& (380	& 44)\,meV	
	& (420	& 90)\,meV
	& \cite{Ait70,Ajz72} \\
1837	& 5464	& $7/2^+$
	& (2.10	& 0.14)\,eV	
	& (2.5	& 0.4)\,eV
	& \cite{Dix71b} \\
1884	& 5501	& $3/2^+$
	& (3.56	& 0.34)\,eV	
	& (4.2	& 1.1)\,eV
	& \cite{Pri57,Ajz72} \\
1927	& 5535	& $5/2^+$
	& (344	& 40)\,meV	
	& (480	& 110)\,meV
	& \cite{Ait70,Ajz72} \\
2036	& 5621	& $5/2^-$
	& (323	& 38)\,meV	
	& (370	& 90)\,meV
	& \cite{Rog72b} \\
2437	& 5938	& $1/2^+$
	& (416	& 48)\,meV	
	& (530	& 130)\,meV
	& \cite{Rog72b} \\
2604	& 6070	& $7/2^+$
	& (2.10	& 0.26)\,eV	
	& (2.7	& 0.54)\,eV
	& \cite{Rog72b} \\
2627	& 6088	& $3/2^-$
	& (5.0	& 0.6)\,eV	
	& (4.5	& 0.9)\,eV
	& \cite{Rog72b} \\
2642	& 6100	& $9/2^-$
	& (440	& 69)\,meV	
	& \multicolumn{2}{c}{$-$}
	& $-$ \\
\end{tabular}
\end{ruledtabular}
\end{table}

\begin{turnpage}
\begin{table*}
\caption{
\label{tab:branch}
Experimentally determined branching ratios (in \%) for 14 resonances of the
reaction 
$^{15}$N($\alpha$,$\gamma$)$^{19}$F are given (first lines) and
compared to literature values (second lines) from
\cite{Til95}. Several new branching ratios have been observed for the
first time. The detection limit for
weak branchings is typically much smaller than 1\,\%.
Spins and parities $J^\pi$ and excitation energies $E_x$ have been
taken from \cite{Til95}. All excitation energies are given in keV. 
}
\begin{ruledtabular}
\begin{tabular}{cr@{$\pm$}lr@{$\pm$}lr@{$\pm$}lr@{$\pm$}lr@{$\pm$}lr@{$\pm$}lr@{$\pm$}lr@{$\pm$}lr@{$\pm$}lr@{$\pm$}lr@{$\pm$}lr@{$\pm$}lr@{$\pm$}lr@{$\pm$}lr@{$\pm$}l}
%
\rule[-2mm]{0mm}{5mm}
$J^\pi$;$E_{x,{\rm{i}}} \rightarrow$
& \multicolumn{2}{c}{0}
& \multicolumn{2}{c}{110}
& \multicolumn{2}{c}{197}
& \multicolumn{2}{c}{1346}
& \multicolumn{2}{c}{1459}
& \multicolumn{2}{c}{1554}
& \multicolumn{2}{c}{2780}
& \multicolumn{2}{c}{3908}
& \multicolumn{2}{c}{3999}
& \multicolumn{2}{c}{4033}
& \multicolumn{2}{c}{4378}
& \multicolumn{2}{c}{4550}
& \multicolumn{2}{c}{4556}
& \multicolumn{2}{c}{4683} 
& \multicolumn{2}{c}{5107} \\
\hline
\rule[-1.5mm]{0mm}{4.5mm}
$5/2^+$;4550
& 4.0 & 2.0
& \multicolumn{2}{c}{$-$}
& 69.8 &  7.2
&  3.5 &  1.9
&  9.7 &  1.8
& 12.9 &  2.0
& \multicolumn{2}{c}{$-$}
& \multicolumn{2}{c}{$-$}
& \multicolumn{2}{c}{$-$}
& \multicolumn{2}{c}{$-$}
& \multicolumn{2}{c}{$-$}
& \multicolumn{2}{c}{$-$}
& \multicolumn{2}{c}{$-$}
& \multicolumn{2}{c}{$-$}
& \multicolumn{2}{c}{$-$} \\
\rule[-1.5mm]{0mm}{4.5mm}
\cite{Til95}
& \multicolumn{2}{c}{$< 5$
  \footnote{See Sect.~\ref{sec:45504556} and \cite{Ajz87}.}}
& \multicolumn{2}{c}{$-$} 
& 69 &  7
&  5 &  3
&  8 &  3
& 18 &  4
& \multicolumn{2}{c}{$-$}
& \multicolumn{2}{c}{$-$}
& \multicolumn{2}{c}{$-$}
& \multicolumn{2}{c}{$-$}
& \multicolumn{2}{c}{$-$}
& \multicolumn{2}{c}{$-$}
& \multicolumn{2}{c}{$-$}
& \multicolumn{2}{c}{$-$}
& \multicolumn{2}{c}{$-$} \\
\hline
\rule[-1.5mm]{0mm}{4.5mm}
$3/2^-$;4556
& \multicolumn{2}{c}{$-$} 
& \multicolumn{2}{c}{$\ast$\footnote{Observed transition; see Sect.~\ref{sec:45504556}.}}
& \multicolumn{2}{c}{$-$}
& \multicolumn{2}{c}{$-$}
& \multicolumn{2}{c}{$-$}
& \multicolumn{2}{c}{$-$}
& \multicolumn{2}{c}{$-$}
& \multicolumn{2}{c}{$-$}
& \multicolumn{2}{c}{$-$}
& \multicolumn{2}{c}{$-$}
& \multicolumn{2}{c}{$-$}
& \multicolumn{2}{c}{$-$}
& \multicolumn{2}{c}{$-$}
& \multicolumn{2}{c}{$-$}
& \multicolumn{2}{c}{$-$} \\
\rule[-1.5mm]{0mm}{4.5mm}
\cite{Til95}
& 36 &  4
& 45 &  5
&  9 &  3
&  4 &  3
& \multicolumn{2}{c}{$< 4$}
&  6 &  3
& \multicolumn{2}{c}{$-$}
& \multicolumn{2}{c}{$-$}
& \multicolumn{2}{c}{$-$}
& \multicolumn{2}{c}{$-$}
& \multicolumn{2}{c}{$-$}
& \multicolumn{2}{c}{$-$}
& \multicolumn{2}{c}{$-$}
& \multicolumn{2}{c}{$-$}
& \multicolumn{2}{c}{$-$} \\
\hline
\rule[-1.5mm]{0mm}{4.5mm}
$5/2^-$;4683
& \multicolumn{2}{c}{$-$}
&  1.3 &  0.5
&  7.7 &  0.7
& 62.8 &  3.0
& 26.9 &  1.4
&  1.4 &  0.3
& \multicolumn{2}{c}{$-$}
& \multicolumn{2}{c}{$-$}
& \multicolumn{2}{c}{$-$}
& \multicolumn{2}{c}{$-$}
& \multicolumn{2}{c}{$-$}
& \multicolumn{2}{c}{$-$}
& \multicolumn{2}{c}{$-$}
& \multicolumn{2}{c}{$-$}
& \multicolumn{2}{c}{$-$} \\
\rule[-1.5mm]{0mm}{4.5mm}
\cite{Til95}
& \multicolumn{2}{c}{$-$}
& \multicolumn{2}{c}{$< 1.5$}
&  5.6 &  0.9
& 63.1 &  3.8
& 31.3 &  2.2
& \multicolumn{2}{c}{$< 5$}
& \multicolumn{2}{c}{$-$}
& \multicolumn{2}{c}{$-$}
& \multicolumn{2}{c}{$-$}
& \multicolumn{2}{c}{$-$}
& \multicolumn{2}{c}{$-$}
& \multicolumn{2}{c}{$-$}
& \multicolumn{2}{c}{$-$}
& \multicolumn{2}{c}{$-$}
& \multicolumn{2}{c}{$-$} \\
\hline
\rule[-1.5mm]{0mm}{4.5mm}
$5/2^+$;5107
& \multicolumn{2}{c}{$-$}
& \multicolumn{2}{c}{$-$}
& 77.8 &  3.8
& \multicolumn{2}{c}{$-$}
&  8.4 &  0.6
&  3.2 &  0.4
&  0.7 &  0.3
& \multicolumn{2}{c}{$6.9^{+0.5}_{-1.0}$}
& \multicolumn{2}{c}{$-$}
& \multicolumn{2}{c}{$-$}
&  3.0 &  0.3
& \multicolumn{2}{c}{$-$}
& \multicolumn{2}{c}{$-$}
& \multicolumn{2}{c}{$-$}
& \multicolumn{2}{c}{$-$} \\
\rule[-1.5mm]{0mm}{4.5mm}
\cite{Til95}
& \multicolumn{2}{c}{$-$}
& \multicolumn{2}{c}{$-$}
& 79.7 &  3.7
& \multicolumn{2}{c}{$< 1.6$}
& 10.4 &  2.7
&  1.8 &  1.8
&  0.7 &  0.6
&  5.4 &  0.9
& \multicolumn{2}{c}{$-$}
& \multicolumn{2}{c}{$-$}
&  2.0 &  0.5
& \multicolumn{2}{c}{$-$}
& \multicolumn{2}{c}{$-$}
& \multicolumn{2}{c}{$-$}
& \multicolumn{2}{c}{$-$} \\
\hline
\rule[-1.5mm]{0mm}{4.5mm}
$1/2^+$;5337
& 38.9 &  1.0
& 40.2 &  1.1
& \multicolumn{2}{c}{$-$}
& \multicolumn{2}{c}{$-$}
& 20.0 &  0.6
&  0.8 &  0.1
& \multicolumn{2}{c}{$-$}
&  0.1 &  0.02
& \multicolumn{2}{c}{$-$}
& \multicolumn{2}{c}{$-$}
& \multicolumn{2}{c}{$-$}
& \multicolumn{2}{c}{$-$}
& \multicolumn{2}{c}{$-$}
& \multicolumn{2}{c}{$-$}
& \multicolumn{2}{c}{$-$} \\
\rule[-1.5mm]{0mm}{4.5mm}
\cite{Til95}
& 37 & 4
& 42 & 4
& \multicolumn{2}{c}{$-$} 
& \multicolumn{2}{c}{$-$} 
& 20 & 2
& \multicolumn{2}{c}{$< 2$}
& \multicolumn{2}{c}{$-$}
& \multicolumn{2}{c}{$-$}
& \multicolumn{2}{c}{$-$}
& \multicolumn{2}{c}{$-$}
& \multicolumn{2}{c}{$-$}
& \multicolumn{2}{c}{$-$}
& \multicolumn{2}{c}{$-$}
& \multicolumn{2}{c}{$-$}
& \multicolumn{2}{c}{$-$} \\
\hline
\rule[-1.5mm]{0mm}{4.5mm}
$7/2^-$;5418
& \multicolumn{2}{c}{$-$}
& \multicolumn{2}{c}{$-$}
& \multicolumn{2}{c}{$-$}
& 72.7 &  2.8
& 12.4 &  0.7
& \multicolumn{2}{c}{$-$}
& \multicolumn{2}{c}{$-$}
& \multicolumn{2}{c}{$-$}
&  9.3 &  0.4
&  5.2 &  0.3
& \multicolumn{2}{c}{$-$}
& \multicolumn{2}{c}{$-$}
& \multicolumn{2}{c}{$-$}
&  0.4 &  0.1
& \multicolumn{2}{c}{$-$} \\
\rule[-1.5mm]{0mm}{4.5mm}
\cite{Til95}
& \multicolumn{2}{c}{$-$} 
& \multicolumn{2}{c}{$-$} 
& \multicolumn{2}{c}{$-$} 
& \multicolumn{2}{c}{70}
& \multicolumn{2}{c}{13}
& \multicolumn{2}{c}{$-$} 
& \multicolumn{2}{c}{$-$} 
& \multicolumn{2}{c}{$-$} 
& \multicolumn{2}{c}{10}
& \multicolumn{2}{c}{6 }
& \multicolumn{2}{c}{$-$}
& \multicolumn{2}{c}{$-$}
& \multicolumn{2}{c}{$-$}
& \multicolumn{2}{c}{$-$}
& \multicolumn{2}{c}{$-$} \\
\hline
\rule[-1.5mm]{0mm}{4.5mm}
$7/2^+$;5464
& \multicolumn{2}{c}{$-$}
& \multicolumn{2}{c}{$-$}
&  3.8 &  0.5
& 32.5 &  1.0
& \multicolumn{2}{c}{$-$}
&  4.9 &  0.2
& 58.1 &  1.7
& \multicolumn{2}{c}{$-$}
& \multicolumn{2}{c}{$-$}
& \multicolumn{2}{c}{$-$}
&  0.2 &  0.03
&  0.5 &  0.03
& \multicolumn{2}{c}{$-$}
& \multicolumn{2}{c}{$-$}
& \multicolumn{2}{c}{$-$} \\
\rule[-1.5mm]{0mm}{4.5mm}
\cite{Til95}
& \multicolumn{2}{c}{$-$} 
& \multicolumn{2}{c}{$-$} 
& \multicolumn{2}{c}{4}
& \multicolumn{2}{c}{32}
& \multicolumn{2}{c}{$-$}
& \multicolumn{2}{c}{5}
& \multicolumn{2}{c}{59}
& \multicolumn{2}{c}{$-$}
& \multicolumn{2}{c}{$-$}
& \multicolumn{2}{c}{$-$}
& \multicolumn{2}{c}{$-$}
& \multicolumn{2}{c}{$-$}
& \multicolumn{2}{c}{$-$}
& \multicolumn{2}{c}{$-$}
& \multicolumn{2}{c}{$-$} \\
\hline
\rule[-1.5mm]{0mm}{4.5mm}
$3/2^+$;5501
& \multicolumn{2}{c}{$-$}
& 21.0 &  0.8
& 49.1 &  1.7
& 16.6 &  0.6
&  1.5 &  0.2
& 11.7 &  0.5
& \multicolumn{2}{c}{$-$}
& \multicolumn{2}{c}{$-$}
& \multicolumn{2}{c}{$-$}
& \multicolumn{2}{c}{$-$}
& \multicolumn{2}{c}{$-$}
& \multicolumn{2}{c}{$-$}
& \multicolumn{2}{c}{$-$}
& \multicolumn{2}{c}{$-$}
& \multicolumn{2}{c}{$-$} \\
\rule[-1.5mm]{0mm}{4.5mm}
\cite{Til95}
& \multicolumn{2}{c}{$-$} 
& \multicolumn{2}{c}{25}
& \multicolumn{2}{c}{49}
& \multicolumn{2}{c}{16}
& \multicolumn{2}{c}{$-$}
& \multicolumn{2}{c}{11}
& \multicolumn{2}{c}{$-$}
& \multicolumn{2}{c}{$-$}
& \multicolumn{2}{c}{$-$}
& \multicolumn{2}{c}{$-$}
& \multicolumn{2}{c}{$-$}
& \multicolumn{2}{c}{$-$}
& \multicolumn{2}{c}{$-$}
& \multicolumn{2}{c}{$-$}
& \multicolumn{2}{c}{$-$} \\
\hline
\rule[-1.5mm]{0mm}{4.5mm}
$5/2^+$;5535
&  6.7 &  0.4
& \multicolumn{2}{c}{$-$}
& 38.6 &  2.3
& \multicolumn{2}{c}{$-$}
& 50.3 &  2.0
&  2.0 &  0.2
& \multicolumn{2}{c}{$-$}
&  0.6 &  0.06
& \multicolumn{2}{c}{$-$}
& \multicolumn{2}{c}{$-$}
&  0.9 &  0.1
& \multicolumn{2}{c}{$-$}
&  0.4 &  0.08
& \multicolumn{2}{c}{$-$}
&  0.5 &  0.09 \\
\rule[-1.5mm]{0mm}{4.5mm}
\cite{Til95}
& \multicolumn{2}{c}{7}
& \multicolumn{2}{c}{$-$} 
& \multicolumn{2}{c}{47}
& \multicolumn{2}{c}{$-$} 
& \multicolumn{2}{c}{45}
& \multicolumn{2}{c}{$-$} 
& \multicolumn{2}{c}{$-$} 
& \multicolumn{2}{c}{$-$} 
& \multicolumn{2}{c}{$-$}
& \multicolumn{2}{c}{$-$}
& \multicolumn{2}{c}{$-$}
& \multicolumn{2}{c}{$-$}
& \multicolumn{2}{c}{$-$}
& \multicolumn{2}{c}{$-$}
& \multicolumn{2}{c}{$-$} \\
\hline
\rule[-1.5mm]{0mm}{4.5mm}
$5/2^-$;5621
& \multicolumn{2}{c}{$-$}
& \multicolumn{2}{c}{$-$}
& 36.8 &  2.0
& 47.8 &  2.5
&  4.6 &  0.4
& \multicolumn{2}{c}{$-$}
& \multicolumn{2}{c}{$-$}
& \multicolumn{2}{c}{$-$}
&  8.0 &  0.5
& \multicolumn{2}{c}{$-$}
& \multicolumn{2}{c}{$-$}
&  1.2 &  0.2
& \multicolumn{2}{c}{$-$}
&  1.5 &  0.2
& \multicolumn{2}{c}{$-$} \\
\rule[-1.5mm]{0mm}{4.5mm}
\cite{Til95}
& \multicolumn{2}{c}{$-$} 
& \multicolumn{2}{c}{$-$} 
& 39 &  4
& 61 &  4
& \multicolumn{2}{c}{$-$}
& \multicolumn{2}{c}{$-$}
& \multicolumn{2}{c}{$-$}
& \multicolumn{2}{c}{$-$}
& \multicolumn{2}{c}{$-$}
& \multicolumn{2}{c}{$-$}
& \multicolumn{2}{c}{$-$}
& \multicolumn{2}{c}{$-$}
& \multicolumn{2}{c}{$-$}
& \multicolumn{2}{c}{$-$}
& \multicolumn{2}{c}{$-$} \\
\hline
\rule[-1.5mm]{0mm}{4.5mm}
$1/2^+$;5938
&  5.3 &  0.3
& 20.6 &  0.9
&  0.9 &  0.2
& \multicolumn{2}{c}{$-$}
& 65.0 &  2.4
&  0.4 &  0.2
& \multicolumn{2}{c}{$-$}
&  7.9 &  0.4
& \multicolumn{2}{c}{$-$}
& \multicolumn{2}{c}{$-$}
& \multicolumn{2}{c}{$-$}
& \multicolumn{2}{c}{$-$}
& \multicolumn{2}{c}{$-$}
& \multicolumn{2}{c}{$-$}
& \multicolumn{2}{c}{$-$} \\
\rule[-1.5mm]{0mm}{4.5mm}
\cite{Til95}
&  7 &  4
& 20 &  6
&  2 &  1
& \multicolumn{2}{c}{$-$}
& 63 &  6
& \multicolumn{2}{c}{$< 2$}
& \multicolumn{2}{c}{$-$}
&  8 &  3
& \multicolumn{2}{c}{$-$}
& \multicolumn{2}{c}{$-$}
& \multicolumn{2}{c}{$-$}
& \multicolumn{2}{c}{$-$}
& \multicolumn{2}{c}{$-$}
& \multicolumn{2}{c}{$-$}
& \multicolumn{2}{c}{$-$} \\
\hline
\rule[-1.5mm]{0mm}{4.5mm}
$7/2^+$;6070
& \multicolumn{2}{c}{$-$}
& \multicolumn{2}{c}{$-$}
& 52.3 &2.0
& 20.6 &  0.8
& \multicolumn{2}{c}{$-$}
&  1.1 &  0.2
& 20.9 &  0.8
& \multicolumn{2}{c}{$-$}
& \multicolumn{2}{c}{$-$}
&  0.1 &  0.04
&  4.0 &  0.2
&  1.0 &  0.05
& \multicolumn{2}{c}{$-$}
& \multicolumn{2}{c}{$-$}
& \multicolumn{2}{c}{$-$} \\
\rule[-1.5mm]{0mm}{4.5mm}
\cite{Til95}
& \multicolumn{2}{c}{$-$} 
& \multicolumn{2}{c}{$-$} 
& 54 &  5
& 19 &  2
& \multicolumn{2}{c}{$-$}
& \multicolumn{2}{c}{$1^{+1}_{-0.5}$}
& 23 &  3
& \multicolumn{2}{c}{$-$}
& \multicolumn{2}{c}{$-$}
& \multicolumn{2}{c}{$< 1$}
&  4 &  1
& \multicolumn{2}{c}{$-$}
& \multicolumn{2}{c}{$-$}
& \multicolumn{2}{c}{$-$}
& \multicolumn{2}{c}{$-$} \\
\hline
\rule[-1.5mm]{0mm}{4.5mm}
$3/2^-$;6088
& 25.1 &  1.0
& 60.3 &  2.2
& 14.6 &  0.6
& \multicolumn{2}{c}{$-$}
& \multicolumn{2}{c}{$-$}
& \multicolumn{2}{c}{$-$}
& \multicolumn{2}{c}{$-$}
& \multicolumn{2}{c}{$-$}
& \multicolumn{2}{c}{$-$}
& \multicolumn{2}{c}{$-$}
& \multicolumn{2}{c}{$-$}
& \multicolumn{2}{c}{$-$}
& \multicolumn{2}{c}{$-$}
& \multicolumn{2}{c}{$-$}
& \multicolumn{2}{c}{$-$} \\
\rule[-1.5mm]{0mm}{4.5mm}
\cite{Til95}
& 25 &  4
& 61 &  5
& 14 &  3
& \multicolumn{2}{c}{$-$}
& \multicolumn{2}{c}{$-$}
& \multicolumn{2}{c}{$-$}
& \multicolumn{2}{c}{$-$}
& \multicolumn{2}{c}{$-$}
& \multicolumn{2}{c}{$-$}
& \multicolumn{2}{c}{$-$}
& \multicolumn{2}{c}{$-$}
& \multicolumn{2}{c}{$-$}
& \multicolumn{2}{c}{$-$}
& \multicolumn{2}{c}{$-$}
& \multicolumn{2}{c}{$-$} \\
\hline
\rule[-1.5mm]{0mm}{4.5mm}
$9/2^-$;6100
& \multicolumn{2}{c}{$-$}
& \multicolumn{2}{c}{$-$}
& 19.4 &  7.2
& 11.4 &  1.6
& \multicolumn{2}{c}{$-$}
& \multicolumn{2}{c}{$< 4.4$}
&  2.9 &  0.8
& \multicolumn{2}{c}{$-$}
& 16.9 &  1.5
& 41.1 &  3.5
&  8.4 &  0.8
& \multicolumn{2}{c}{$-$}
& \multicolumn{2}{c}{$-$}
& \multicolumn{2}{c}{$-$}
& \multicolumn{2}{c}{$-$} \\
\rule[-1.5mm]{0mm}{4.5mm}
\cite{Til95}
& \multicolumn{2}{c}{$-$} 
& \multicolumn{2}{c}{$-$}
& \multicolumn{2}{c}{$-$}
& \multicolumn{2}{c}{$-$}
& \multicolumn{2}{c}{$-$}
& \multicolumn{2}{c}{$-$}
& \multicolumn{2}{c}{$-$}
& \multicolumn{2}{c}{$-$}
& \multicolumn{2}{c}{$-$}
& \multicolumn{2}{c}{$-$}
& \multicolumn{2}{c}{$-$}
& \multicolumn{2}{c}{$-$}
& \multicolumn{2}{c}{$-$}
& \multicolumn{2}{c}{$-$}
& \multicolumn{2}{c}{$-$} \\
\end{tabular}
\end{ruledtabular}
\end{table*}
\end{turnpage}

\subsection{
\label{sec:403343784648}
${\bm{J^\pi = 9/2^-}}$, ${\bm{E_x = 4033}}$\,keV and
${\bm{J^\pi = 7/2^+}}$, ${\bm{E_x = 4378}}$\,keV and
${\bm{J^\pi = 13/2^+}}$, ${\bm{E_x = 4648}}$\,keV
} 
No attempt was made to measure these three resonances in this
experiment. The resonances at 4033\,keV and 4648\,keV do not
significantly contribute to the astrophysical reaction rate. Their
strengths are very small because the partial widths $\Gamma_\alpha$ are
suppressed as a consequence of the large centrifugal barrier.

Nevertheless,
the astrophysical reaction rate around $T_9 = 0.1 - 0.2$ is strongly
influenced by the resonance at $E_x = 4378$\,keV. The resonance
strength of this $7/2^+$ resonance is entirely determined by its 
partial width $\Gamma_\alpha$ which is confirmed by the measured ratio
$\Gamma_\gamma/\Gamma > 0.96$ \cite{Pri89}. The partial width
$\Gamma_\alpha$ has been measured in the
$^{15}$N($^7$Li,t)$^{19}$F reaction: $\Gamma_\alpha =
1.5^{+1.5}_{-0.8}$\,neV \cite{Oli96,Oli97}. This leads to a resonance
strength of $\omega 
\gamma = 6^{+6}_{-3}$\,neV which is far below our experimental
sensitivity of the order of 1\,$\mu$eV.

\subsection{
\label{sec:45504556}
${\bm{J^\pi = 5/2^+}}$, ${\bm{E_x = 4550}}$\,keV and
${\bm{J^\pi = 3/2^-}}$, ${\bm{E_x = 4556}}$\,keV} 
Both resonances have been discussed in detail in \cite{Wil95}; the
experimental thick-target yield is shown in Fig.~2 of \cite{Wil95}. For the
lower resonance a weak branch to the ground state has been detected
for the first time in the ($\alpha$,$\gamma$) reaction. With the
assumption $\Gamma_\alpha \ll \Gamma_\gamma \approx \Gamma$ and the
total width $\Gamma = 101 \pm 55$\,meV from a lifetime measurement
\cite{Kiss82} one obtains a ground state radiation width
$\Gamma_{\gamma_0} \approx 4$\,meV in excellent agreement with the
transition strength in \cite{Endt79}: $B(E2) = 1.0 \pm 0.2$\,W.u.\
corresponding to $\Gamma_{\gamma_0} = 4.8 \pm 1.0$\,meV. The total
strength of $\omega \gamma = 95.5 \pm 11.7$\,$\mu$eV is in
excellent agreement with a previous result \cite{Mag87,Til95}.

For the very weak resonance at $E_x = 4556$\,keV only the strongest
branching to the first excited state at $E_x = 110$\,keV could be
detected. Together with the known branching ratio of 45\,\%
\cite{Til95} a resonance strength of $\omega \gamma = 6.4 \pm
2.5$\,$\mu$eV was derived. This value agrees with the previously
adopted upper limit of
$\omega \gamma \le 10$\,$\mu$eV \cite{Mag87}.

\subsection{
\label{sec:4683}
${\bm{J^\pi = 5/2^-}}$, ${\bm{E_x = 4683}}$\,keV} 
The total strength of this resonance is $\omega \gamma = 5.6 \pm
0.6$\,meV which is in good agreement with the adopted strength of
$\omega \gamma = 6 \pm 1$\,meV \cite{Rog72a,Til95}. Two weak branching
ratios to the states at $E_x = 110$ and 1554\,keV have been detected
for the first time; both branching ratios are close to the upper
limits in \cite{Til95}.

\subsection{
\label{sec:5107}
${\bm{J^\pi = 5/2^+}}$, ${\bm{E_x = 5107}}$\,keV} 
The total strength of this resonance is $\omega \gamma = 9.7 \pm
1.6$\,meV. The uncertainties have been reduced significantly compared
to the adopted value of 
$\omega \gamma = 13 \pm 8$\,meV \cite{Ait70,Til95}. Six branching
ratios for this resonance have been confirmed with slightly improved
accuracies. 

From the measured ratio $\Gamma_\gamma / \Gamma = 0.97 \pm 0.03$
\cite{Pri89} and our resonance strength one can deduce $\Gamma_\alpha
= 3.3 \pm 0.6$\,meV. The measured upper limit of the lifetime leads to
a total width $\Gamma > 22$\,meV \cite{Til95}, and thus $\Gamma_\gamma
> 19$\,meV for this state.

\subsection{
\label{sec:5337}
${\bm{J^\pi = 1/2^{(+)}}}$, ${\bm{E_x = 5337}}$\,keV} 
Spin and parity $J^\pi = 1/2^{(+)}$ have been adopted in
\cite{Til95}. As we shall show later, there is clear confirmation of
the positive parity from our experimental data (see
Sect.~\ref{sec:clu}). Therefore, the resonance will be labelled by
$J^\pi = 1/2^+$ in the following discussion.

This resonance allows a precise comparison of this gas target
experiment to previous experiments which were performed using solid
targets. The previously adopted strength of this resonance of $\omega
\gamma = 1.64 \pm 0.16$\,eV \cite{Dix71a,Til95} has a small
uncertainty of about 10\,\%. The resonance is relatively strong and
has been seen in several later experiments. Often, this resonance has
been used as calibration standard for other resonances (see discussion
in \cite{Ajz72}). Because of
$J^\pi = 1/2^+$ the angular distribution for all decay $\gamma$-rays
is isotropic. This leads to small uncertainties for the branching
ratios. Within the experimental uncertainties, we confirm the adopted
branching ratios \cite{Til95}. Additionally, two weak decay branchings
to states at $E_x = 1554$ and 3908\,keV have been observed. Our
observed resonance strength $\omega \gamma = 1.69 \pm 0.14$\,eV is
3\,\% higher than the adopted value, but agrees nicely within the
uncertainties. 

The total width $\Gamma$ could be determined from the ana\-lysis of the
measured yield curve over a broad energy interval of about 100\,keV
following the procedure outlined in \cite{Koe99}. The result of $\Gamma
= 1.3 \pm 0.5$\,keV replaces the adopted lower limit
$\Gamma > 6.6$\,eV. For this resonance $\Gamma_\alpha \approx \Gamma$.
This fact is confirmed by
the observation of this $1/2^+$ resonance in the elastic scattering
data (see Fig.~\ref{fig:elast}). Thus, $\Gamma_\gamma = 1.69 \pm
0.14$\,eV.

\subsection{
\label{sec:54185464}
${\bm{J^\pi = 7/2^-}}$, ${\bm{E_x = 5418}}$\,keV and
${\bm{J^\pi = 7/2^+}}$, ${\bm{E_x = 5464}}$\,keV} 
Because of the relative large $J = 7/2$ of both resonances, transitions
to the ground state and first excited state of $^{19}$F with $J = 1/2$
are very unlikely. Indeed, no resonant enhancement can be seen in the
yield curves. The observed yields are small and can be completely
understood as the tails of the broad $3/2^+$ resonance at $E_x =
5501$\,keV (see Sect.~\ref{sec:5501} and \ref{sec:clu}). A resonant
enhancement was found for 5 transitions in the 5418\,keV resonance
(including one new weak branching) and for 6 transitions in the
5464\,keV resonance (including two new weak branchings). The summed
strengths are $\omega \gamma = 380 \pm 44$\,meV for the 5418\,keV
resonance and $\omega \gamma = 2.10 \pm 0.14$\,eV for the 5464\,keV
resonance. For both strengths and branching ratios good agreement with
the adopted values \cite{Til95,Ait70,Ajz72,Dix71b} is
found. No uncertainties are given for the adopted
branching ratios \cite{Til95}.

For the 5418\,keV resonance the ratio $\Gamma_\gamma / \Gamma = 0.04
\pm 0.007$ has been measured \cite{Pri89} which allows to calculate
the partial widths $\Gamma_\alpha = 2.4 \pm 0.3$\,eV and
$\Gamma_\gamma = 98 \pm 12$\,meV. This leads to a total width $\Gamma
= 2.5 \pm 0.4$\,eV which agrees perfectly with the adopted $\Gamma =
2.6 \pm 0.7$\,eV. 

For the 5464\,keV resonance only an upper limit
$\Gamma_\gamma / \Gamma < 0.028$ was measured \cite{Pri89} which leads
to the partial widths $\Gamma_\alpha > 18.8$\,eV and $\Gamma_\gamma
\approx 525 \pm 35$\,meV. For the total width we obtain $\Gamma >
19.3$\,eV, which is not in 
contradiction with a lifetime measurement of $\Gamma > 2.5$\,eV.

\subsection{
\label{sec:5501}
${\bm{J^\pi = 3/2^+}}$, ${\bm{E_x = 5501}}$\,keV} 
The strength of this relatively broad resonance is $\omega \gamma =
3.56 \pm 0.34$\,eV which is slightly smaller than the adopted value of
$4.2 \pm 1.1$\,eV \cite{Pri57,Ajz72,Til95}. A correction of 11\,\% was
taken into account because of the deviation of the elastic scattering
cross section from Rutherford in this resonance (see
Fig.~\ref{fig:elast} and Sect.~\ref{sec:clu}). Tails of this resonance
could be observed over a broad interval of about 400\,keV, and from
the yield curve a total width of $\Gamma = 4.7 \pm 1.6$\,keV could be
derived in the same way as for the 5337\,keV resonance (see
Sect.~\ref{sec:5337}). This valus agrees nicely with the adopted width
$\Gamma = 4 \pm 1$\,keV \cite{Til95}.

With the exception of the transition to the first excited state at
$E_x = 110$\,keV, all branching ratios agree with the adopted values
\cite{Til95} which are again given without uncertainties. One new
branching has been measured to $E_x = 1459$\,keV. 

From the resonance strength one can directly calculate the radiation
width $\Gamma_\gamma = 1.78 \pm 0.17$\,eV.
Because of the large total width one can assume $\Gamma_\alpha \approx
\Gamma$ and therefore $\Gamma_\alpha \gg \Gamma_\gamma$ for this
resonance. This assumption is comfirmed by the elastic scattering data
(see Sect.~\ref{sec:clu}).

\subsection{
\label{sec:55355621}
${\bm{J^\pi = 5/2^+}}$, ${\bm{E_x = 5535}}$\,keV and
${\bm{J^\pi = 5/2^-}}$, ${\bm{E_x = 5621}}$\,keV
} 
Both $J = 5/2$ resonances lie in the high-energy tail of the broad
5501\,keV resonance. Therefore, for several transitions to low-lying
states in $^{19}$F a small contribution from the braod 5501\,keV
resonance has to be subtracted. The summed strengths are $\omega
\gamma = 344 \pm 40$\,meV for the 5535\,keV resonance and $\omega
\gamma = 323 \pm 38$\,meV for the 5621\,keV resonance. Both values are
in agreement with the adopted values within their large uncertainties.

Eight branching ratios (including five new ones) have been measured
for the 5535\,keV resonance which are in reasonable agreement with the
adopted values \cite{Til95}. For the 5621\,keV resonance six
branchings (including four new ones) were measured.

\subsection{
\label{sec:5938}
${\bm{J^\pi = 1/2^+}}$, ${\bm{E_x = 5938}}$\,keV} 
The resonance at $E_x = 5938$\,keV is more than 400\,keV above the
broad 5501\,keV resonance and 150\,keV below the broad 6088\,keV
resonance thus not influenced by their tails. We find a total strength
of $\omega \gamma = 416 \pm 48$\,meV slightly lower than the adopted
value of $\omega \gamma = 530 \pm 130$\,meV \cite{Rog72b,Til95}.
Six branching ratios were determined including a new weak $0.4 \pm
0.2$\,\% branching to the $E_x = 1554$\,keV state.

\subsection{
\label{sec:607060886100}
${\bm{J^\pi = 3/2^-}}$, ${\bm{E_x = 6070}}$\,keV and
${\bm{J^\pi = 7/2^+}}$, ${\bm{E_x = 6088}}$\,keV and
${\bm{J^\pi = 9/2^-}}$, ${\bm{E_x = 6100}}$\,keV
} 
These three resonances are located close to each other, and the
central 6088\,keV resonance has a rather large width of $\Gamma =
4$\,keV. This value is confirmed by the analysis of our yield curve
which leads to $\Gamma = 4.7 \pm 1.6$\,keV. The lower resonace is a
factor of four narrower: $\Gamma = 1.2$\,keV, and the width is unknown
for the 6100\,keV resonance.

From a careful analysis of the yield curves including Doppler shifts
in the $\gamma$ spectra (see also \cite{Wil95}) we are able to
disentangle the contributions of these three resonances. Fortunately,
these resonances show very different branching ratios which made the
analysis less complicated than expected. The total strengths are
$\omega \gamma = 2.1 \pm 0.26$\,eV, $5.0 \pm 0.6$\,eV, and $440 \pm
69$\,meV for $E_x = 6070$\,keV, 6088\,keV, and 6100\,keV. The first
two numbers are in good agreement with previous data
\cite{Rog72b,Til95} whereas the resonance at $E_x = 6100$\,keV has
been measured for the first time in the
$^{15}$N($\alpha$,$\gamma$)$^{19}$F reaction.
The measured branching ratios agree perfectly with the adopted values
for the $E_x = 6088$\,keV resonance; two new branchings have been
detected in the $E_x = 6070$\,keV resonance, and no branching ratios
are given in \cite{Til95} for the $E_x = 6100$\,keV resonance.

\section{\label{sec:clu}$\bm\alpha$-cluster states in $^{19}$F and
direct capture}
$\alpha$ clustering is a well-known phenomenon in light
nuclei. Characteristic properties of states with strong $\alpha$
clustering are large $\alpha$-particle spectroscopic factors which
can be measured in $\alpha$ transfer reactions and relatively large
partial widths $\Gamma_\alpha$ for states above threshold. Many
properties of $^{19}$F can be understood from such a simple model
which neglects excitations in the $\alpha$ particle and the $^{15}$N
core. The ground state of the nucleus $^{15}$N can be described as
$^{16}$O with a hole in the $p_{1/2}$ shell.

The cluster wave function $u_{NLJ}(r)$ which describes the relative
motion of the $\alpha$ particle and the $^{15}$N core is characterized
by the node number $N$ and the orbital and total angular momenta $L$
and $J$ with $J = L \pm 1/2$. The cluster $N$ and $L$ values are
related to the corresponding quantum numbers $n_i$ and $l_i$ of the
four nucleons forming the $\alpha$ cluster in the $sd$ shell:
\begin{equation}
Q = 2N + L = \sum_{i=1}^{4}(2n_i + l_i) = \sum_{i=1}^{4} q_i
\label{eq:q}
\end{equation}

For $^{20}$Ne one expects five cluster states with positive parity for
the ground state band ($Q = 8$, $J^\pi = 0^+, 2^+, 4^+, 6^+, 8^+$) and
five cluster states with negative parity ($Q = 9$, $J^\pi = 1^-, 3^-,
5^-, 7^-, 9^-$). The coupling of these bands with the ground state
spin of $^{15}$N ($J^\pi = 1/2^-$) leads to close doublets in $^{19}$F
because of a weak spin-orbit coupling (with the exception of the $L =
0$, $J^\pi = 1/2^-$ state which cannot be split into a doublet).

The properties of these bands have been discussed in detail in
\cite{Abe93,Buck77,Des87,Duf00} (and references therein). 
In the following we shall focus our interest on
the $Q = 9$, $L = 1$, $J^\pi = 1/2^+, 3/2^+$ doublet which appears as
resonances in the $^{15}$N($\alpha$,$\gamma$)$^{19}$F reaction at $E_x
= 5337$\,keV and $5501$\,keV.

Following the procedure outlined in \cite{Abe93,Mohr94} we tried to
describe these resonances as so-called potential resonances. A full
discussion of potential resonances is also given in \cite{Mohr98}. 
Our analysis uses systematic folding potentials which have been shown
to describe many properties of the systems $^{15}$N $\otimes$ $\alpha$
and $^{16}$O $\otimes$ $\alpha$ \cite{Abe93} including a weak
spin-orbit potential with a shape $\sim 1/r \, dV/dr$. The strength of
the central potential has been adjusted to reproduce the properly
weighted average energy of the $L = 1$ doublet, and the strength of
the spin-orbit potential was adjusted to the energy difference between
both states. A similar procedure was performed for $L = $ even partial
waves. One finds similar normalization parameters of the folding
potential of $\lambda \approx 1.3$ for even and odd partial waves.
After this adjustment the potential is fixed, and the elastic
scattering wave function can be calculated without any further
adjustment. One finds a good agreement with the experimentally
measured excitation functions of elastic scattering
which were already shown in
Fig.~\ref{fig:elast}. However, the widths of both resonances are
slightly overestimated. For the $1/2^+$ state at $E_x = 5337$\,keV we
find $\Gamma_\alpha^{\rm{calc}} = 3.44$\,keV compared to
$\Gamma_\alpha^{\rm{exp}} = 1.3 \pm 0.5$\,keV (from this work), and
for the $3/2^+$ state at $E_x = 5501$\,keV we find
$\Gamma_\alpha^{\rm{calc}} = 9.99$\,keV compared to
$\Gamma_\alpha^{\rm{exp}} = 4.2 \pm 0.9$\,keV (weighted average of the
adopted value and this work). In both cases $\Gamma_\alpha^{\rm{exp}}
\approx 0.4 \times \Gamma_\alpha^{\rm{calc}}$ which shows that the
wave functions of both resonances consist of about 40\,\% $\alpha$
cluster contribution. Reduced widths have been calculated for these
states leading to $\theta_\alpha^2 = 0.53$ and $0.52$ for the $E_x =
5337$\,keV state from the generator coordinate method (GCM)
\cite{Des87} and from the orthogonality condition model (OCM)
\cite{Sak79} and to $\theta_\alpha^2 = 0.46$ and $0.50$ for the $E_x =
5501$\,keV state. A similar result is also obtained for the $L = 1$,
$J^\pi = 1^-$ resonance in $^{20}$Ne (see Fig.~10 of \cite{Abe93}).

\begin{figure}[hbt]
\includegraphics[ bb = 92 84 506 337, width = 85 mm, clip]{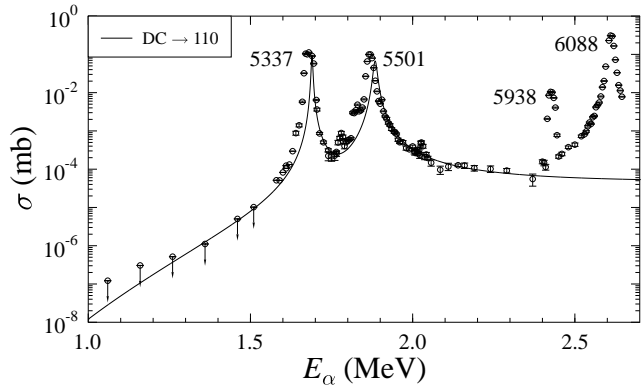}
\caption{
\label{fig:res} 
The calculated direct capture cross section for the transition to the
first excited state with $J^\pi = 1/2^-$ at $E_x = 110$\,keV is
compared to experimental 
data. One finds good agreement between experiment and theory in the
$J^\pi = 1/2^+$ and $3/2^+$ resonances 
at $E_x = 5337$\,keV and 5501\,keV
and in a broad energy interval
at lower and higher energies.
}
\end{figure}

From the wave functions of the potential model the direct capture
(DC) cross section can be 
calculated in the $E_x = 5337$\,keV and $5501$\,keV resonances and at
lower and higher energies. The full formalism has been described in
\cite{Mohr94}. The 
calculated DC cross section for the transition to the $E_x
= 110$\,keV state is compared to experimental data in
Fig.~\ref{fig:res}. The spectroscopic factor of the final state has
been taken from \cite{Sak79}. Note that the $E_x = 110$\,keV state is
the lowest $\alpha$ cluster state in $^{19}$F with $Q = 8$, $N =4$,
$L=0$, and $J^\pi = 1/2^-$. Good agreement is found in the
resonances and in a broad energy interval at lower and higher
energies. Similar results have been obtained for the transitions to
the $\alpha$ cluster states with $L = 2$ and $J^\pi = 5/2^-$ at $E_x =
1346$\,keV and $J^\pi = 3/2^-$ at $E_x = 1459$\,keV. Especially, the
relatively small cross section of the transition $J^\pi = 3/2^+$, $E_x
= 5501$\,keV $\rightarrow$ $J^\pi = 3/2^-$, $E_x = 1459$\,keV (see
Table \ref{tab:branch}) is also
reproduced by the DC calculations.

If the energy differs sufficiently from the resonance
energy, the calculated cross section depends only weakly on the
energy. However, this ``direct capture'' results from the tail of the
$E_x = 5337$\,keV and $5501$\,keV resonances; the experimental proof
is the identical 
branching ratio which is observed in the $E_x = 5501$\,keV resonance
and in the whole energy range from 5400\,keV $\alt E_x \alt$ 5700\,keV
(of course, excluding the narrow resonances at $E_x = 5418$\,keV,
5464\,keV, 5535\,keV, and 5621\,keV).

\section{\label{sec:ast}Astrophysical Reaction Rate}
The astrophysical reaction rate of $^{15}$N($\alpha$,$\gamma$)$^{19}$F
is dominated by resonant contributions. In total, 48 resonances have
been taken into account in the NACRE compilation \cite{NACRE}. For
the most important low-lying resonances the strengths from a transfer
experiment \cite{Oli96} were adopted, and our results from a first
analysis of our experimental data \cite{Wil97} were taken into
account. In cases where several determinations of the resonance
strength were available the weighted average has been adopted which is
close to our values because of our small experimental
uncertainties. Because the resonance strengths in this work are
identical to \cite{Wil97} the adopted reaction rate of \cite{NACRE}
does not change.

\begin{figure}[hbt]
\includegraphics[ bb = 92 84 509 337, width = 85 mm, clip]{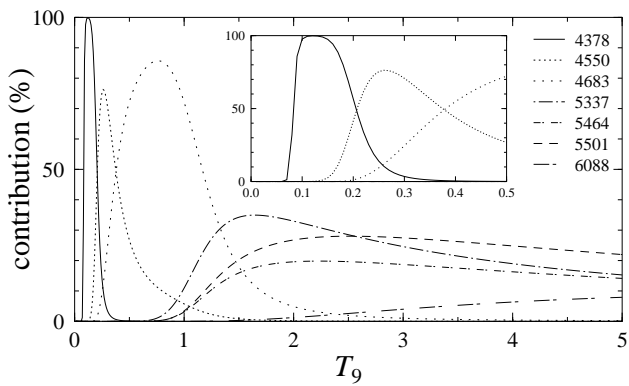}
\caption{
\label{fig:rateratio} 
Contribution of seven individual resonances to the total reaction rate 
$N_A <\sigma v>$ of
the reaction $^{15}$N($\alpha$,$\gamma$)$^{19}$F in per cent. The
inset shows the low temperature region ($T_9 \le 0.5$). Below
$T_9 \alt 1.2$ the reaction rate is dominated by three individual
resonances at $E_x = 4378$, 4550, and 4683\,keV. At higher
temperatures the reaction rate is given by the 
sum of several strong resonances.
}
\end{figure}

The influence of the various resonances to the astrophysical reaction
rate is shown in Fig.~\ref{fig:rateratio}.
 The total reaction rate has been taken from the fit function provided
by the NACRE compilation \cite{NACRE}. The contribution of each
resonance has been calculated using the strengths from Table
\ref{tab:strength} and the formalism for narrow
resonances.

Below $T_9 = 0.08$ direct capture becomes relevant \cite{Oli96}; however, the
absolute reaction rate at $T_9 = 0.08$ drops below $N_A <\sigma v> \, = \,
10^{-25}$\,cm$^3$\,s$^{-1}$\,mole$^{-1}$ which is considered as
negligible in usual astrophysical calculations \cite{NACRE}.

At $0.1 \le T_9 \le 0.2$ the reaction rate is completely determined by
the $7/2^+$ state at 4378\,keV. Therefore the uncertainty of the
reaction rate is given by the uncertainty of this resonance strength
which is a factor of two \cite{Oli96}. 

The uncertainty of the reaction rate is significantly reduced at
higher temperatures by the reduced experimental uncertainties of this
work. In the temperature interval $0.2 \alt T_9 \alt 0.4$ the $5/2^+$
resonance at 4550\,keV is dominating, and the interval $0.4 \alt T_9
\alt 1.2$ is 
governed by the $5/2^-$ resonance at 4683\,keV. Above $T_9 \agt 1.2$
several strong resonances are contributing to the reaction rate with
typical contributions of the order of 10\,\% $-$ 30\,\%. The
experimental uncertainties for these resonance strengths of about
10\,\% lead to an experimentally determined reaction rate with similar
uncertainty \cite{NACRE}. Because of the relatively high excitation
energy of the first excited state in $^{15}$N, the reaction rate which
is calculated from laboratory data, is identical to the reaction rate
under stellar conditions \cite{NACRE}.

\section{\label{sec:con}Conclusions}
We have measured 14 resonances of the reaction
$^{15}$N($\alpha$,$\gamma$)$^{19}$F at low energies between 0.6\,MeV
$\le E_\alpha \le$ 2.7\,MeV. The combination of the {\sc{dynamitron}}
high-current accelerator, the windowless gas target {\sc{rhinoceros}},
and large HPGe detectors with 
active shieldings leads to the excellent sensitivity of the experimental
set-up.

Two resonances have been detected for the first time in the
($\alpha$,$\gamma$) channel. The uncertainties of the resonance
strengths are reduced significantly. Branching ratios could be
determined with improved precision; roughly 20 new weak branchings
were seen for the first time. Typical detection limits are of the
order of 1\,$\mu$eV for resonance strengths and far below 1\,\% for weak
branching ratios. Additionally, partial widths $\Gamma_\alpha$ and
$\Gamma_\gamma$ were derived from the
combination of experimental resonance strengths and measured ratios
$\Gamma_\gamma/\Gamma$ \cite{Pri89}. 

The states at $E_x = 5337$\,keV and 5501\,keV have been identified
unambiguously as the $L = 1$, $J^\pi = 1/2^+$ and $3/2^+$ doublet of
$^{15}$N $\otimes$ $\alpha$ cluster states. This identification is
based on the large partial widths $\Gamma_\alpha$, the angular
distribution of elastic scattering, and the decay branchings of these
states. The analysis of the yield curves over a broad energy interval has
allowed to determine the total width of the $E_x = 5337$\,keV state
for the first time and to confirm the total width of the $E_x =
5501$\,keV state. For both levels we find reduced widths
$\theta_\alpha^2 \approx 0.4$ in good agreement with various
theoretical predictions.

The astrophysical reaction rate of the NACRE compilation \cite{NACRE}
remains unchanged. The resonance strengths in this work are identical
to our previous analysis \cite{Wil97} which has been taken into
account in \cite{NACRE}. The uncertainties of the reaction rate have
been reduced significantly at temperatures above $T_9 = 0.2$.

\begin{acknowledgments}
We thank U.~Kneissl for supporting this experiment. The help of
R.~Kunz, A.~Mayer, and the {\sc{dynamitron}} crew during the beamtimes
is gratefully acknowledged. This work was 
supported by Deutsche Forschungsgemeinschaft DFG (Sta290).
\end{acknowledgments}

\bibliography{n15a}

\begin{thebibliography}{37}
\expandafter\ifx\csname natexlab\endcsname\relax\def\natexlab#1{#1}\fi
\expandafter\ifx\csname bibnamefont\endcsname\relax
  \def\bibnamefont#1{#1}\fi
\expandafter\ifx\csname bibfnamefont\endcsname\relax
  \def\bibfnamefont#1{#1}\fi
\expandafter\ifx\csname citenamefont\endcsname\relax
  \def\citenamefont#1{#1}\fi
\expandafter\ifx\csname url\endcsname\relax
  \def\url#1{\texttt{#1}}\fi
\expandafter\ifx\csname urlprefix\endcsname\relax\def\urlprefix{URL }\fi
\providecommand{\bibinfo}[2]{#2}
\providecommand{\eprint}[2][]{\url{#2}}

\bibitem[{\citenamefont{{Forestini} et~al.}(1992)\citenamefont{{Forestini},
  {Goriely}, {Jorissen}, and {Arnould}}}]{1992A&A...261..157F}
\bibinfo{author}{\bibfnamefont{M.}~\bibnamefont{{Forestini}}},
  \bibinfo{author}{\bibfnamefont{S.}~\bibnamefont{{Goriely}}},
  \bibinfo{author}{\bibfnamefont{A.}~\bibnamefont{{Jorissen}}},
  \bibnamefont{and}
  \bibinfo{author}{\bibfnamefont{M.}~\bibnamefont{{Arnould}}},
  \bibinfo{journal}{Astron.\ Astroph.} \textbf{\bibinfo{volume}{261}},
  \bibinfo{pages}{157} (\bibinfo{year}{1992}).

\bibitem[{\citenamefont{{Jorissen} et~al.}(1992)\citenamefont{{Jorissen},
  {Smith}, and {Lambert}}}]{1992A&A...261..164J}
\bibinfo{author}{\bibfnamefont{A.}~\bibnamefont{{Jorissen}}},
  \bibinfo{author}{\bibfnamefont{V.~V.} \bibnamefont{{Smith}}},
  \bibnamefont{and} \bibinfo{author}{\bibfnamefont{D.~L.}
  \bibnamefont{{Lambert}}}, \bibinfo{journal}{Astron.\ Astroph.}
  \textbf{\bibinfo{volume}{261}}, \bibinfo{pages}{164} (\bibinfo{year}{1992}).

\bibitem[{\citenamefont{{Meynet} and {Arnould}}(2000)}]{2000A&A...355..176M}
\bibinfo{author}{\bibfnamefont{G.}~\bibnamefont{{Meynet}}} \bibnamefont{and}
  \bibinfo{author}{\bibfnamefont{M.}~\bibnamefont{{Arnould}}},
  \bibinfo{journal}{Astron.\ Astroph.} \textbf{\bibinfo{volume}{355}},
  \bibinfo{pages}{176} (\bibinfo{year}{2000}).

\bibitem[{\citenamefont{{Woosley} et~al.}(1990)\citenamefont{{Woosley},
  {Hartmann}, {Hoffman}, and {Haxton}}}]{1990ApJ...356..272W}
\bibinfo{author}{\bibfnamefont{S.~E.} \bibnamefont{{Woosley}}},
  \bibinfo{author}{\bibfnamefont{D.~H.} \bibnamefont{{Hartmann}}},
  \bibinfo{author}{\bibfnamefont{R.~D.} \bibnamefont{{Hoffman}}},
  \bibnamefont{and} \bibinfo{author}{\bibfnamefont{W.~C.}
  \bibnamefont{{Haxton}}}, \bibinfo{journal}{\apj}
  \textbf{\bibinfo{volume}{356}}, \bibinfo{pages}{272} (\bibinfo{year}{1990}).

\bibitem[{\citenamefont{{{G{\"o}rres}, J.}}(2002)}]{Gor2002}
\bibinfo{author}{\bibnamefont{{{G{\"o}rres}, J.}}}, in
  \emph{\bibinfo{booktitle}{Nuclei in the Cosmos VII}}, edited by
  \bibinfo{editor}{\bibnamefont{{S.\ Kubono {\it et al.}}}}
  (\bibinfo{year}{2002}).

\bibitem[{\citenamefont{{Wallace} and {Woosley}}(1981)}]{1981ApJS...45..389W}
\bibinfo{author}{\bibfnamefont{R.~K.} \bibnamefont{{Wallace}}}
  \bibnamefont{and} \bibinfo{author}{\bibfnamefont{S.~E.}
  \bibnamefont{{Woosley}}}, \bibinfo{journal}{\apj Suppl.}
  \textbf{\bibinfo{volume}{45}}, \bibinfo{pages}{389} (\bibinfo{year}{1981}).

\bibitem[{\citenamefont{{Langanke} et~al.}(1986)\citenamefont{{Langanke},
  {Wiescher}, {Fowler}, and {G{\"o}rres}}}]{1986ApJ...301..629L}
\bibinfo{author}{\bibfnamefont{K.}~\bibnamefont{{Langanke}}},
  \bibinfo{author}{\bibfnamefont{M.}~\bibnamefont{{Wiescher}}},
  \bibinfo{author}{\bibfnamefont{W.~A.} \bibnamefont{{Fowler}}},
  \bibnamefont{and}
  \bibinfo{author}{\bibfnamefont{J.}~\bibnamefont{{G{\"o}rres}}},
  \bibinfo{journal}{\apj} \textbf{\bibinfo{volume}{301}}, \bibinfo{pages}{629}
  (\bibinfo{year}{1986}).

\bibitem[{\citenamefont{Butt et~al.}(1998)\citenamefont{Butt, Hammer, Jaeger,
  Kunz, Mayer, Parker, Schreiter, and Staudt}}]{Butt98}
\bibinfo{author}{\bibfnamefont{Y.~M.} \bibnamefont{Butt}},
  \bibinfo{author}{\bibfnamefont{J.~W.} \bibnamefont{Hammer}},
  \bibinfo{author}{\bibfnamefont{M.}~\bibnamefont{Jaeger}},
  \bibinfo{author}{\bibfnamefont{R.}~\bibnamefont{Kunz}},
  \bibinfo{author}{\bibfnamefont{A.}~\bibnamefont{Mayer}},
  \bibinfo{author}{\bibfnamefont{P.~D.} \bibnamefont{Parker}},
  \bibinfo{author}{\bibfnamefont{R.}~\bibnamefont{Schreiter}},
  \bibnamefont{and} \bibinfo{author}{\bibfnamefont{G.}~\bibnamefont{Staudt}},
  \bibinfo{journal}{\prc} \textbf{\bibinfo{volume}{58}}, \bibinfo{pages}{R10}
  (\bibinfo{year}{1998}).

\bibitem[{\citenamefont{deOliveira et~al.}(1997)\citenamefont{deOliveira, Coc,
  Aguer, Bogaert, Kiener, Lefebvre, Tatischeff, Thibaud, Fortier, Maison
  et~al.}}]{Oli97}
\bibinfo{author}{\bibfnamefont{F.}~\bibnamefont{deOliveira}},
  \bibinfo{author}{\bibfnamefont{A.}~\bibnamefont{Coc}},
  \bibinfo{author}{\bibfnamefont{P.}~\bibnamefont{Aguer}},
  \bibinfo{author}{\bibfnamefont{G.}~\bibnamefont{Bogaert}},
  \bibinfo{author}{\bibfnamefont{J.}~\bibnamefont{Kiener}},
  \bibinfo{author}{\bibfnamefont{A.}~\bibnamefont{Lefebvre}},
  \bibinfo{author}{\bibfnamefont{V.}~\bibnamefont{Tatischeff}},
  \bibinfo{author}{\bibfnamefont{J.-P.} \bibnamefont{Thibaud}},
  \bibinfo{author}{\bibfnamefont{S.}~\bibnamefont{Fortier}},
  \bibinfo{author}{\bibfnamefont{J.~M.} \bibnamefont{Maison}},
  \bibnamefont{et~al.}, \bibinfo{journal}{\prc} \textbf{\bibinfo{volume}{55}},
  \bibinfo{pages}{3149} (\bibinfo{year}{1997}).

\bibitem[{\citenamefont{Magnus et~al.}(1987)\citenamefont{Magnus, Smith,
  Parker, Azuma, Campbell, King, and Vise}}]{Mag87}
\bibinfo{author}{\bibfnamefont{P.~V.} \bibnamefont{Magnus}},
  \bibinfo{author}{\bibfnamefont{M.~S.} \bibnamefont{Smith}},
  \bibinfo{author}{\bibfnamefont{P.~D.} \bibnamefont{Parker}},
  \bibinfo{author}{\bibfnamefont{R.~E.} \bibnamefont{Azuma}},
  \bibinfo{author}{\bibfnamefont{C.}~\bibnamefont{Campbell}},
  \bibinfo{author}{\bibfnamefont{J.~D.} \bibnamefont{King}}, \bibnamefont{and}
  \bibinfo{author}{\bibfnamefont{J.}~\bibnamefont{Vise}},
  \bibinfo{journal}{Nucl.\ Phys.\ A} \textbf{\bibinfo{volume}{470}},
  \bibinfo{pages}{206} (\bibinfo{year}{1987}).

\bibitem[{\citenamefont{Rogers et~al.}(1972{\natexlab{a}})\citenamefont{Rogers,
  Aitken, and Litherland}}]{Rog72a}
\bibinfo{author}{\bibfnamefont{D.~W.~O.} \bibnamefont{Rogers}},
  \bibinfo{author}{\bibfnamefont{J.~H.} \bibnamefont{Aitken}},
  \bibnamefont{and} \bibinfo{author}{\bibfnamefont{A.~E.}
  \bibnamefont{Litherland}}, \bibinfo{journal}{Can.\ J.\ Phys.}
  \textbf{\bibinfo{volume}{50}}, \bibinfo{pages}{268}
  (\bibinfo{year}{1972}{\natexlab{a}}).

\bibitem[{\citenamefont{Rogers et~al.}(1972{\natexlab{b}})\citenamefont{Rogers,
  Beukens, and Diamond}}]{Rog72b}
\bibinfo{author}{\bibfnamefont{D.~W.~O.} \bibnamefont{Rogers}},
  \bibinfo{author}{\bibfnamefont{R.~P.} \bibnamefont{Beukens}},
  \bibnamefont{and} \bibinfo{author}{\bibfnamefont{W.~T.}
  \bibnamefont{Diamond}}, \bibinfo{journal}{Can.\ J.\ Phys.}
  \textbf{\bibinfo{volume}{50}}, \bibinfo{pages}{2428}
  (\bibinfo{year}{1972}{\natexlab{b}}).

\bibitem[{\citenamefont{Dixon and Storey}(1971)}]{Dix71a}
\bibinfo{author}{\bibfnamefont{W.~R.} \bibnamefont{Dixon}} \bibnamefont{and}
  \bibinfo{author}{\bibfnamefont{R.~S.} \bibnamefont{Storey}},
  \bibinfo{journal}{Can.\ J.\ Phys.} \textbf{\bibinfo{volume}{49}},
  \bibinfo{pages}{1714} (\bibinfo{year}{1971}).

\bibitem[{\citenamefont{Dixon et~al.}(1971)\citenamefont{Dixon, Storey, Aitken,
  Litherland, and Rogers}}]{Dix71b}
\bibinfo{author}{\bibfnamefont{W.~R.} \bibnamefont{Dixon}},
  \bibinfo{author}{\bibfnamefont{R.~S.} \bibnamefont{Storey}},
  \bibinfo{author}{\bibfnamefont{J.~H.} \bibnamefont{Aitken}},
  \bibinfo{author}{\bibfnamefont{A.~E.} \bibnamefont{Litherland}},
  \bibnamefont{and} \bibinfo{author}{\bibfnamefont{D.~W.~O.}
  \bibnamefont{Rogers}}, \bibinfo{journal}{\prl} \textbf{\bibinfo{volume}{27}},
  \bibinfo{pages}{1460} (\bibinfo{year}{1971}).

\bibitem[{\citenamefont{Aitken et~al.}(1970)\citenamefont{Aitken, Azuma,
  Litherland, Charlesworth, Rogers, and Simpson}}]{Ait70}
\bibinfo{author}{\bibfnamefont{J.~H.} \bibnamefont{Aitken}},
  \bibinfo{author}{\bibfnamefont{R.~E.} \bibnamefont{Azuma}},
  \bibinfo{author}{\bibfnamefont{A.~E.} \bibnamefont{Litherland}},
  \bibinfo{author}{\bibfnamefont{A.~M.} \bibnamefont{Charlesworth}},
  \bibinfo{author}{\bibfnamefont{D.~W.~O.} \bibnamefont{Rogers}},
  \bibnamefont{and} \bibinfo{author}{\bibfnamefont{J.~J.}
  \bibnamefont{Simpson}}, \bibinfo{journal}{Can.\ J.\ Phys.}
  \textbf{\bibinfo{volume}{48}}, \bibinfo{pages}{1617} (\bibinfo{year}{1970}).

\bibitem[{\citenamefont{Price}(1957)}]{Pri57}
\bibinfo{author}{\bibfnamefont{P.~C.} \bibnamefont{Price}},
  \bibinfo{journal}{Proc.\ Phys.\ Soc.\ (London) A}
  \textbf{\bibinfo{volume}{70}}, \bibinfo{pages}{661} (\bibinfo{year}{1957}).

\bibitem[{\citenamefont{Ajzenberg-Selove}(1987)}]{Ajz87}
\bibinfo{author}{\bibfnamefont{F.}~\bibnamefont{Ajzenberg-Selove}},
  \bibinfo{journal}{Nucl.\ Phys.\ A} \textbf{\bibinfo{volume}{475}},
  \bibinfo{pages}{1} (\bibinfo{year}{1987}).

\bibitem[{\citenamefont{Tilley et~al.}(1995)\citenamefont{Tilley, Weller,
  Cheves, and Chasteler}}]{Til95}
\bibinfo{author}{\bibfnamefont{D.~R.} \bibnamefont{Tilley}},
  \bibinfo{author}{\bibfnamefont{H.~R.} \bibnamefont{Weller}},
  \bibinfo{author}{\bibfnamefont{C.~M.} \bibnamefont{Cheves}},
  \bibnamefont{and} \bibinfo{author}{\bibfnamefont{R.~M.}
  \bibnamefont{Chasteler}}, \bibinfo{journal}{Nucl.\ Phys.\ A}
  \textbf{\bibinfo{volume}{595}}, \bibinfo{pages}{1} (\bibinfo{year}{1995}).

\bibitem[{\citenamefont{Angulo et~al.}(1999)\citenamefont{Angulo, Arnould,
  Rayet, Descouvemont, Baye, Leclercq-Willain, Coc, Barhoumi, Aguer, Rolfs
  et~al.}}]{NACRE}
\bibinfo{author}{\bibfnamefont{C.}~\bibnamefont{Angulo}},
  \bibinfo{author}{\bibfnamefont{M.}~\bibnamefont{Arnould}},
  \bibinfo{author}{\bibfnamefont{M.}~\bibnamefont{Rayet}},
  \bibinfo{author}{\bibfnamefont{P.}~\bibnamefont{Descouvemont}},
  \bibinfo{author}{\bibfnamefont{D.}~\bibnamefont{Baye}},
  \bibinfo{author}{\bibfnamefont{C.}~\bibnamefont{Leclercq-Willain}},
  \bibinfo{author}{\bibfnamefont{A.}~\bibnamefont{Coc}},
  \bibinfo{author}{\bibfnamefont{S.}~\bibnamefont{Barhoumi}},
  \bibinfo{author}{\bibfnamefont{P.}~\bibnamefont{Aguer}},
  \bibinfo{author}{\bibfnamefont{C.}~\bibnamefont{Rolfs}},
  \bibnamefont{et~al.}, \bibinfo{journal}{Nucl.\ Phys.\ A}
  \textbf{\bibinfo{volume}{656}}, \bibinfo{pages}{1} (\bibinfo{year}{1999}).

\bibitem[{\citenamefont{{de Oliveira} et~al.}(1996)\citenamefont{{de Oliveira},
  {Coc}, {Aguer}, {Angulo}, {Bogaert}, {Kiener}, {Lefebvre}, {Tatischeff},
  {Thibaud}, {Fortier} et~al.}}]{Oli96}
\bibinfo{author}{\bibfnamefont{F.}~\bibnamefont{{de Oliveira}}},
  \bibinfo{author}{\bibfnamefont{A.}~\bibnamefont{{Coc}}},
  \bibinfo{author}{\bibfnamefont{P.}~\bibnamefont{{Aguer}}},
  \bibinfo{author}{\bibfnamefont{C.}~\bibnamefont{{Angulo}}},
  \bibinfo{author}{\bibfnamefont{G.}~\bibnamefont{{Bogaert}}},
  \bibinfo{author}{\bibfnamefont{J.}~\bibnamefont{{Kiener}}},
  \bibinfo{author}{\bibfnamefont{A.}~\bibnamefont{{Lefebvre}}},
  \bibinfo{author}{\bibfnamefont{V.}~\bibnamefont{{Tatischeff}}},
  \bibinfo{author}{\bibfnamefont{J.-P.} \bibnamefont{{Thibaud}}},
  \bibinfo{author}{\bibfnamefont{S.}~\bibnamefont{{Fortier}}},
  \bibnamefont{et~al.}, \bibinfo{journal}{Nucl.\ Phys.\ A}
  \textbf{\bibinfo{volume}{597}}, \bibinfo{pages}{231} (\bibinfo{year}{1996}).

\bibitem[{\citenamefont{Hammer}(1999)}]{Hammer99}
\bibinfo{author}{\bibfnamefont{J.~W.} \bibnamefont{Hammer}},
  \bibinfo{type}{Tech. Rep.}, \bibinfo{institution}{Institut f\"ur
  Strahlenphysik, Universit\"at Stuttgart} (\bibinfo{year}{1999}).

\bibitem[{\citenamefont{K{\"o}lle et~al.}(1999)\citenamefont{K{\"o}lle,
  K{\"o}lle, Braitmayr, Mohr, Wilmes, Staudt, Hammer, Jaeger, Knee, Kunz
  et~al.}}]{Koe99}
\bibinfo{author}{\bibfnamefont{V.}~\bibnamefont{K{\"o}lle}},
  \bibinfo{author}{\bibfnamefont{U.}~\bibnamefont{K{\"o}lle}},
  \bibinfo{author}{\bibfnamefont{S.~E.} \bibnamefont{Braitmayr}},
  \bibinfo{author}{\bibfnamefont{P.}~\bibnamefont{Mohr}},
  \bibinfo{author}{\bibfnamefont{S.}~\bibnamefont{Wilmes}},
  \bibinfo{author}{\bibfnamefont{G.}~\bibnamefont{Staudt}},
  \bibinfo{author}{\bibfnamefont{J.~W.} \bibnamefont{Hammer}},
  \bibinfo{author}{\bibfnamefont{M.}~\bibnamefont{Jaeger}},
  \bibinfo{author}{\bibfnamefont{H.}~\bibnamefont{Knee}},
  \bibinfo{author}{\bibfnamefont{R.}~\bibnamefont{Kunz}}, \bibnamefont{et~al.},
  \bibinfo{journal}{Nucl.\ Inst.\ Meth.\ Phys.\ Res.\ A}
  \textbf{\bibinfo{volume}{431}}, \bibinfo{pages}{160} (\bibinfo{year}{1999}).

\bibitem[{\citenamefont{Wilmes}(1996)}]{Wilmes96}
\bibinfo{author}{\bibfnamefont{S.}~\bibnamefont{Wilmes}}, Ph.D. thesis,
  \bibinfo{school}{Universit\"at T\"ubingen} (\bibinfo{year}{1996}).

\bibitem[{\citenamefont{Pringle and Vermeer}(1989)}]{Pri89}
\bibinfo{author}{\bibfnamefont{D.~M.} \bibnamefont{Pringle}} \bibnamefont{and}
  \bibinfo{author}{\bibfnamefont{W.~J.} \bibnamefont{Vermeer}},
  \bibinfo{journal}{Nucl.\ Phys.\ A} \textbf{\bibinfo{volume}{499}},
  \bibinfo{pages}{117} (\bibinfo{year}{1989}).

\bibitem[{\citenamefont{Ziegler}(1977)}]{Zie77}
\bibinfo{author}{\bibfnamefont{J.~F.} \bibnamefont{Ziegler}},
  \emph{\bibinfo{title}{Helium Stopping Powers and Ranges in All Elements}}
  (\bibinfo{publisher}{Pergamon Press}, \bibinfo{address}{New York},
  \bibinfo{year}{1977}).

\bibitem[{\citenamefont{Ajzenberg-Selove}(1972)}]{Ajz72}
\bibinfo{author}{\bibfnamefont{F.}~\bibnamefont{Ajzenberg-Selove}},
  \bibinfo{journal}{Nucl.\ Phys.\ A} \textbf{\bibinfo{volume}{190}},
  \bibinfo{pages}{1} (\bibinfo{year}{1972}).

\bibitem[{\citenamefont{Wilmes et~al.}(1995)\citenamefont{Wilmes, Mohr,
  Atzrott, K{\"o}lle, Staudt, Mayer, and Hammer}}]{Wil95}
\bibinfo{author}{\bibfnamefont{S.}~\bibnamefont{Wilmes}},
  \bibinfo{author}{\bibfnamefont{P.}~\bibnamefont{Mohr}},
  \bibinfo{author}{\bibfnamefont{U.}~\bibnamefont{Atzrott}},
  \bibinfo{author}{\bibfnamefont{V.}~\bibnamefont{K{\"o}lle}},
  \bibinfo{author}{\bibfnamefont{G.}~\bibnamefont{Staudt}},
  \bibinfo{author}{\bibfnamefont{A.}~\bibnamefont{Mayer}}, \bibnamefont{and}
  \bibinfo{author}{\bibfnamefont{J.~W.} \bibnamefont{Hammer}},
  \bibinfo{journal}{\prc} \textbf{\bibinfo{volume}{52}}, \bibinfo{pages}{R2823}
  (\bibinfo{year}{1995}).

\bibitem[{\citenamefont{Kiss et~al.}(1982)\citenamefont{Kiss, Nyak{\'o},
  Somorjai, Antilla, and Bister}}]{Kiss82}
\bibinfo{author}{\bibfnamefont{{\'A}.~Z.} \bibnamefont{Kiss}},
  \bibinfo{author}{\bibfnamefont{B.}~\bibnamefont{Nyak{\'o}}},
  \bibinfo{author}{\bibfnamefont{E.}~\bibnamefont{Somorjai}},
  \bibinfo{author}{\bibfnamefont{A.}~\bibnamefont{Antilla}}, \bibnamefont{and}
  \bibinfo{author}{\bibfnamefont{M.}~\bibnamefont{Bister}},
  \bibinfo{journal}{Nucl.\ Inst.\ Meth.\ Phys.\ Res.\ A}
  \textbf{\bibinfo{volume}{203}}, \bibinfo{pages}{107} (\bibinfo{year}{1982}).

\bibitem[{\citenamefont{Endt}(1979)}]{Endt79}
\bibinfo{author}{\bibfnamefont{P.~M.} \bibnamefont{Endt}},
  \bibinfo{journal}{At.\ Data Nucl.\ Data Tables}
  \textbf{\bibinfo{volume}{23}}, \bibinfo{pages}{3} (\bibinfo{year}{1979}).

\bibitem[{\citenamefont{Abele and Staudt}(1993)}]{Abe93}
\bibinfo{author}{\bibfnamefont{H.}~\bibnamefont{Abele}} \bibnamefont{and}
  \bibinfo{author}{\bibfnamefont{G.}~\bibnamefont{Staudt}},
  \bibinfo{journal}{\prc} \textbf{\bibinfo{volume}{47}}, \bibinfo{pages}{742}
  (\bibinfo{year}{1993}).

\bibitem[{\citenamefont{Buck and Pilt}(1977)}]{Buck77}
\bibinfo{author}{\bibfnamefont{B.}~\bibnamefont{Buck}} \bibnamefont{and}
  \bibinfo{author}{\bibfnamefont{A.~A.} \bibnamefont{Pilt}},
  \bibinfo{journal}{Nucl.\ Phys.\ A} \textbf{\bibinfo{volume}{280}},
  \bibinfo{pages}{133} (\bibinfo{year}{1977}).

\bibitem[{\citenamefont{Descouvemont and Baye}(1987)}]{Des87}
\bibinfo{author}{\bibfnamefont{P.}~\bibnamefont{Descouvemont}}
  \bibnamefont{and} \bibinfo{author}{\bibfnamefont{D.}~\bibnamefont{Baye}},
  \bibinfo{journal}{Nucl.~Phys.~A} \textbf{\bibinfo{volume}{463}},
  \bibinfo{pages}{629} (\bibinfo{year}{1987}).

\bibitem[{\citenamefont{Dufour and Descouvemont}(2000)}]{Duf00}
\bibinfo{author}{\bibfnamefont{M.}~\bibnamefont{Dufour}} \bibnamefont{and}
  \bibinfo{author}{\bibfnamefont{P.}~\bibnamefont{Descouvemont}},
  \bibinfo{journal}{Nucl.~Phys.~A} \textbf{\bibinfo{volume}{672}},
  \bibinfo{pages}{153} (\bibinfo{year}{2000}).

\bibitem[{\citenamefont{Mohr et~al.}(1994)\citenamefont{Mohr, K{\"o}lle,
  Wilmes, Atzrott, Staudt, Hammer, Krauss, and Oberhummer}}]{Mohr94}
\bibinfo{author}{\bibfnamefont{P.}~\bibnamefont{Mohr}},
  \bibinfo{author}{\bibfnamefont{V.}~\bibnamefont{K{\"o}lle}},
  \bibinfo{author}{\bibfnamefont{S.}~\bibnamefont{Wilmes}},
  \bibinfo{author}{\bibfnamefont{U.}~\bibnamefont{Atzrott}},
  \bibinfo{author}{\bibfnamefont{G.}~\bibnamefont{Staudt}},
  \bibinfo{author}{\bibfnamefont{J.~W.} \bibnamefont{Hammer}},
  \bibinfo{author}{\bibfnamefont{H.}~\bibnamefont{Krauss}}, \bibnamefont{and}
  \bibinfo{author}{\bibfnamefont{H.}~\bibnamefont{Oberhummer}},
  \bibinfo{journal}{\prc} \textbf{\bibinfo{volume}{50}}, \bibinfo{pages}{1543}
  (\bibinfo{year}{1994}).

\bibitem[{\citenamefont{Mohr et~al.}(1998)\citenamefont{Mohr, Beer, Oberhummer,
  and Staudt}}]{Mohr98}
\bibinfo{author}{\bibfnamefont{P.}~\bibnamefont{Mohr}},
  \bibinfo{author}{\bibfnamefont{H.}~\bibnamefont{Beer}},
  \bibinfo{author}{\bibfnamefont{H.}~\bibnamefont{Oberhummer}},
  \bibnamefont{and} \bibinfo{author}{\bibfnamefont{G.}~\bibnamefont{Staudt}},
  \bibinfo{journal}{\prc} \textbf{\bibinfo{volume}{58}}, \bibinfo{pages}{932}
  (\bibinfo{year}{1998}).

\bibitem[{\citenamefont{Sakuda and Nemoto}(1979)}]{Sak79}
\bibinfo{author}{\bibfnamefont{T.}~\bibnamefont{Sakuda}} \bibnamefont{and}
  \bibinfo{author}{\bibfnamefont{F.}~\bibnamefont{Nemoto}},
  \bibinfo{journal}{Prog.~Theor.~Phys.} \textbf{\bibinfo{volume}{62}},
  \bibinfo{pages}{1606} (\bibinfo{year}{1979}).

\bibitem[{\citenamefont{Wilmes et~al.}(1997)\citenamefont{Wilmes, K{\"o}lle,
  K{\"o}lle, Staudt, Mohr, Hammer, and Mayer}}]{Wil97}
\bibinfo{author}{\bibfnamefont{S.}~\bibnamefont{Wilmes}},
  \bibinfo{author}{\bibfnamefont{V.}~\bibnamefont{K{\"o}lle}},
  \bibinfo{author}{\bibfnamefont{U.}~\bibnamefont{K{\"o}lle}},
  \bibinfo{author}{\bibfnamefont{G.}~\bibnamefont{Staudt}},
  \bibinfo{author}{\bibfnamefont{P.}~\bibnamefont{Mohr}},
  \bibinfo{author}{\bibfnamefont{J.~W.} \bibnamefont{Hammer}},
  \bibnamefont{and} \bibinfo{author}{\bibfnamefont{A.}~\bibnamefont{Mayer}},
  \bibinfo{journal}{Nucl.~Phys.~A} \textbf{\bibinfo{volume}{621}},
  \bibinfo{pages}{145c} (\bibinfo{year}{1997}).

\end{thebibliography}

\end{document}